\documentclass[11pt,twoside]{article}

\usepackage{asp2014}

\aspSuppressVolSlug
\resetcounters

\begin{document}

\title{Provenance Tools for Astronomy}

\author{Mich\`{e}le Sanguillon,$^1$ Fran\c{c}ois Bonnarel,$^2$ Mireille Louys,$^2$$^,$$^3$ Markus Nullmeier,$^4$ Kristin Riebe,$^5$ and Mathieu Servillat$^6$
\affil{$^1$Laboratoire Univers et Particules de Montpellier, Universit\'{e} de Montpellier, CNRS/IN2P3, France; \email{Michele.Sanguillon@umontpellier.fr}}
\affil{$^2$Centre de Donn\'{e}es astronomiques de Strasbourg, Observatoire Astronomique de Strasbourg, Universit\'{e} de Strasbourg, CNRS, Strasbourg, France}
\affil{$^3$ICube Laboratory, Universit\'{e} de Strasbourg, CNRS, Strasbourg, France}
\affil{$^4$Zentrum f{\"u}r Astronomie der Universit{\"a}t Heidelberg, Astronomisches Rechen-Institut, Heidelberg, Germany}
\affil{$^5$Leibniz Institute for Astrophysics Potsdam, Germany}
\affil{$^6$Laboratoire Univers et Th\'{e}ories, Observatoire de Paris, PSL Research University, CNRS, 92190 Meudon, France}}


\paperauthor{Mich\`{e}le Sanguillon}{Michele.Sanguillon@umontpellier.fr}{0000-0003-0196-6301}{CNRS, Universit\'{e} de Montpellier}{LUPM}{Montpellier}{ }{34095}{France}
\paperauthor{Fran\c{c}ois Bonnarel}{francois.bonnarel@astro.unistra.fr}{ }{Universit\'{e} de Strasbourg, CNRS}{Observatoire astronomique de Strasbourg - UMR7550}{Strasbourg}{}{67000}{France}
\paperauthor{Mireille Louys}{mireille.louys@unistra.fr}{0000-0002-4334-1142}{Universit\'{e} de Strasbourg, CNRS} {ICube Laboratory - UMR7357}{Strasbourg}{}{67000}{France}
\paperauthor{Markus Nullmeier}{mnullmei@ari.uni-heidelberg.de}{0000-0003-2807-8961}{Zentrum f{\"u}r Astronomie der Universit{\"a}t Heidelberg}{Astronomisches Rechen-Institut}{Heidelberg}{ }{ }{Germany}
\paperauthor{Kristin Riebe}{kriebe@aip.de}{0000-0001-7161-8869}{Leibniz Institute for Astrophysics Potsdam }{ }{Potsdam}{ }{14482}{Germany}
\paperauthor{Mathieu Servillat}{mathieu.servillat@obspm.fr}{0000-0001-5443-4128}{Observatoire de Paris, PSL Research University, CNRS}{LUTH}{Meudon}{}{92195}{France}

\begin{abstract}
 In the context of astronomy projects, scientists have been confronted with the problem of describing in a standardized way how their data have been produced. 

 As presented in a talk at last year's ADASS, the International Virtual Observatory Alliance (IVOA) is working on the definition of a Provenance Data Model, 
 compatible with the W3C PROV model, which shall describe how provenance metadata can be modeled, stored and exchanged in astronomy.

In this poster, we present the current status of our developments of libraries and tools, mainly open source, which implement the IVOA Provenance Data Model in order to
produce, serve, load and visualize provenance information. These implementations are also needed to validate and adjust the data model and the standard
definitions for accessing provenance. The provenance tools developed and created for the W3C framework are reused and extended when possible to
tackle the domain of astronomical data.
\end{abstract}


\section{Introduction}
The International Virtual Observatory Alliance\footnote{\url{http://www.ivoa.net/}} 
has developed several data models to foster interoperability
between diverse astronomy projects. Even though a lot of objects (spectra, images, simulations, etc.) are already well described, 
some parts of the information about how datasets have been produced is still missing.

That is why the IVOA Data Model Working Group investigates how to model
provenance information of a dataset, how this information can be stored and how it can be exchanged.
In order to check the validity of the defined model, the group implemented the IVOA Provenance Data Model in four environments:
Pollux, CTA, RAVE, and one at CDS.

Here, we present the tools developed to implement this model in these different contexts.
 
\section{IVOA Provenance Data Model}
The IVOA Provenance Data Model
\citep{IVOAProvenanceDM} follows the W3C Provenance definition,
i.~e., that provenance is ``information about entities, activities, and people involved in
producing a piece of data or thing, which can be used to form assessments about its quality, reliability or trustworthiness''.

The main core classes (\emph{Entity}, \emph{Activity}, \emph{Agent}) and its relations (\emph{wasGeneratedBy}, etc.) have the same name as in the W3C Provenance Data Model \citep{std:W3CProvDM}.
We add the \emph{ActivityFlow} class and the \emph{hadStep} relation in order to allow users to describe workflows of activities. 
We also add the possibility to separate the description of an activity or entity from the activity/entity itself.

\articlefigure[width=.7\textwidth]{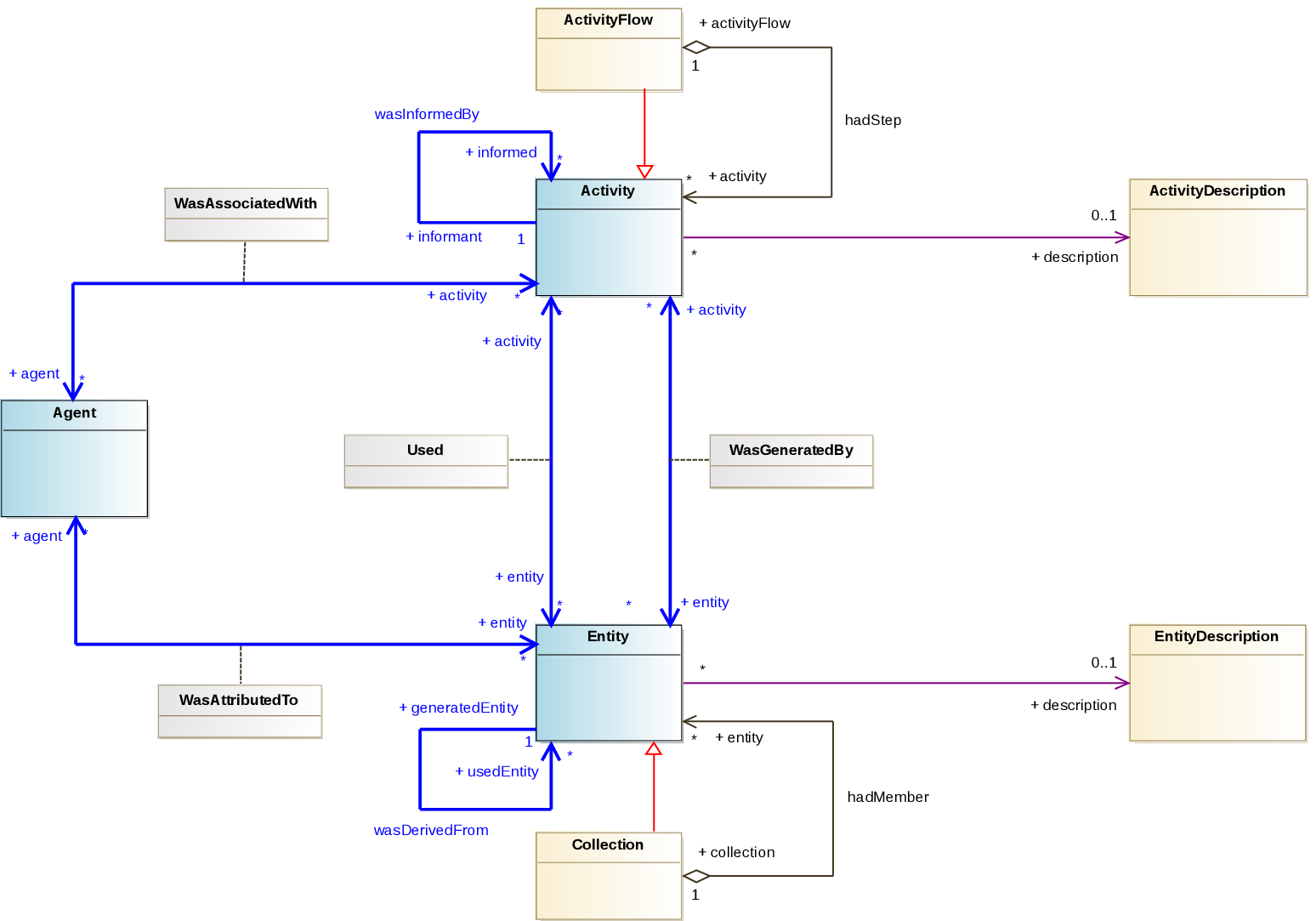}{P9-129_f1}{IVOA Provenance Data Model}

\section{voprov library}
The voprov\footnote{\url{https://github.com/sanguillon/voprov/}} package 
is an open source Python library derived from the prov\footnote{\url{https://github.com/trungdong/prov/}}
Python library (MIT license) developed by Trung Dong Huynh (University
of Southampton).


The voprov package implements the serialization of the IVOA Provenance Data Model. 
As this model is very close to the W3C one, the voprov library uses the following facilites from prov:
the PROV-N, PROV-JSON, and PROV-XML serialization formats, as well as PDF, PNG, and SVG graphical representations.
It adds these IVOA features: 
flows of activities (pipelines), which are composed of different activity
steps, and serialization into the VOTable format.

This library is currently used in the context of the POLLUX database,
which offers high resolution synthetic
spectra computed using the best available models of the atmosphere and efficient
spectral synthesis codes.


When a spectrum is integrated into the database, provenance information is retrieved and serialized 
in different formats and with different levels of detail. 
When a user or a program queries the Pollux database (via the SSA protocol of the Virtual Observatory),
he is informed (via the DataLink protocol) of the existence of a service that allows him to retrieve 
provenance information in a given format and for a given detail level. 
This functionality has been implemented in the CASSIS spectrum visualization tool.

\section{Django package}
The django-prov\_vo package\footnote{\url{https://github.com/kristinriebe/django-prov_vo}} is an open source Python package that can be reused in Django web applications for
serving provenance information. The data model classes are directly mapped to tables
in a relational database. The package provides different interfaces to extract provenance: a REST interface to retrieve lists of entities, activities and agents, and a ProvDAL interface, which is defined in the current IVOA Provenance Working Draft. The ProvDAL interface takes the identifier of an entity, activity or an agent as a parameter and then returns the available provenance information in one of the serialization formats (currently PROV-N and PROV-JSON). A few visualization techniques for the retrieved provenance graph are also included.

This \texttt{django-prov\_vo} package was developed for a provenance service of the RAVE\footnote{\url{https://www.rave-survey.org/}} project.
Within the RAVE (RAdial Velocity Experiment) survey, spectra of about half a million stars from the southern hemisphere were observed and stellar properties determined.

\section{Prototype PostgreSQL database at CDS}
We implemented the IVOA Provenance DM in a test Postgres database at CDS.
The database handles a small collection of image datasets, such as Schmidt plates, mono-band and color composed images or HiPS representations of pixel data. 
From the IVOA Provenance Datamodel specification we designed a database schema and implemented the various related tables recommended in the data model as Postgres tables. 


A small set of plates, with their digitization, cutout extractions,  RGB color composition, and HiPS generation activities, is used to populate the database.
Various scenarios for querying and displaying their provenance information have been tested in SQL.
For query responses, PROV-N, PROV-JSON, and PROV-VOTable formats are provided.
A simple Python API allowing users to select the main types of requests and to display the responses via W3C Prov library has been designed.
It allows users querying for various combinations of provenance relationships in the database and to visualize the provenance graph in a user friendly representation. 

This provides experience with the DM implementation and clues to build up a TAP SCHEMA representation for ProvTAP services, a preliminary version of which has been developed.

\section{UWS Server at Observatoire de Paris}

In the context of the Cherenkov Telescope Array\footnote{\url{https://www.cta-observatory.org/}} (CTA) project, a job control system based on the IVOA UWS pattern has been developed as an open source Python application: OPUS\footnote{\url{https://github.com/mservillat/OPUS}} (Observatoire de Paris UWS System). 
This system has been used to test the execution of CTA data analysis tools on a work cluster. It implements the ProvenanceDM concept of ActivityDescription files and provides the provenance information for each executed job in
PROV-JSON and PROV-XML serializations.

The CTA is the next generation ground-based very high energy gamma-ray instrument. Contrary to previous Cherenkov experiments, it will serve as an open observatory providing data to a wide astrophysics community, with the requirement to offer self-described data products to users that may be unaware of the Cherenkov astronomy specificities (see also \citet{P75_adassxxvii}).

\acknowledgements This work was partially funded by the Federal Ministry of Education and Research in Germany 
and by the ASTERICS project (\url{http://www.asterics2020.eu/}).
Additional funding was provided by the INSU (Action Sp\'{e}cifique Observatoire Virtuel, ASOV), the Grand-Sud-Ouest Data Centre, 
the Paris Astronomical Data Centre, and the Observatoire Astronomique de Strasbourg.

\bibliography{P9-129}  

\end{document}